\documentclass[10pt, aps, prb, preprint, longbibliography, superscriptaddress]{revtex4-2}

\usepackage{amsmath, amssymb}
\usepackage[dvipdfmx]{graphicx}
\usepackage[colorlinks,citecolor=blue,urlcolor=blue,linkcolor=blue]{hyperref}
\usepackage{siunitx}

\newcommand{\figref}[1]{FIG.~\ref{#1}}
\let\orgeqref=\eqref
\renewcommand{\eqref}[1]{Eq.~\orgeqref{#1}}

\newcommand{\ave}[1]{\langle #1 \rangle}

\begin{document}

\title{Analysis of Brownian Motion by Elementary School Students}

\author{Makito Miyazaki}
\affiliation{Hakubi Center for Advanced Research, Kyoto University, Yoshida-honmachi, Sakyo-ku, Kyoto 606-8501, Japan.}
\affiliation{JST PRESTO, 4-1-8, Honcho, Kawaguchi-shi, Saitama 332-0012, Japan.}
\affiliation{Institut Curie, PSL Research University, CNRS, UMR 144, F-75005 Paris, France.}
\author{Yosuke Yamazaki}
\affiliation{Department of Physics, Faculty of Science and Engineering, Waseda University, 3-4-1 Okubo, Shinjuku-ku, Tokyo 169-8555, Japan.}
\author{Yamato Hasegawa}
\affiliation{Tokyo Tech High School of Science and Technology, 3-3-6, Shibaura, Minato-ku, Tokyo 108-0023, Japan.}

\begin{abstract}
	To stimulate the intellectual curiosity of elementary school students,
	we conducted a workshop in distance education aimed at exploring the microscopic world inside a cell.
	In this workshop, elementary school students motivated to learn more on the subject of science
	analyzed movies of the Brownian motion of micrometer-sized particles suspended in water,
	\cite{howard_random, paul_2003, marco_2013}
	using the open-source software, Tracker.\cite{tracker, douglas_2011}
	These students then performed two-dimensional(2D)-random walk experiments
	using a dice game sheet to examine the physical mechanism of Brownian motion.\cite{oosawa_hand}
	After the workshop, we conducted a questionnaire-based survey.
	Many participants answered that the contents were difficult but interesting,
	suggesting that our workshop was very efficient in stimulating the curiosity of motivated students.

\end{abstract}

\maketitle

\section{Introduction}
\label{intro}

In Japan, the curricula for primary and secondary school education are stipulated
in the Course of Study.\cite{shido_elem}
According to the Trends in International Mathematics and Science Study in 2019 (TIMSS2019),
published by the International Association for the Evaluation of Educational Achievement (IEA),\cite{timss}
92\% of elementary school students in Japan answered in the affirmatively
to the item ``I like learning science'', a number that exceeds the international average of 86\%.
However, from the perspective of gifted education,
it is unclear whether motivated students are fascinated with the contents of science classes
stipulated in the \textit{Course of Study}.

Here, we conducted a three-hour workshop for motivated elementary school students entitled
``Let's explore the small world inside a cell!'' twice during the Japanese academic year of 2020.
The workshop was composed of three sections:
In the first section, each participant performed an analysis of the motion of Brownian particles
captured by an optical microscope (lecture: 30 minutes, practice: 40 minutes).
In the second section, the students performed a 2D-random walk experiment
using a dice game sheet to examine the physical mechanism of Brownian motion (lecture: 30 minutes, practice: 20 minutes).
In the last section, the students considered protein dynamics in a cell based on
what they learned in the first and second sections,
then learn the frontiers of life science research and protein engineering (lecture: 30 minutes).
We distributed flyers for the workshop to public elementary schools,
and then recruited 27 motivated students in the 4th to 6th grades
(18 and 9 participants in the first and second workshops, respectively).
Owing to the COVID-19 pandemic, we were forced to provide distance education.
Therefore, we held remote workshops via the Internet using the Zoom application.
YH, a physics teacher at a high school, organized the workshop,
YY, a PhD student in the field of biophysics, developed the data aggregation and analysis tools,
and MM, a principal investigator in the field of biophysics, gave a lecture.

In this workshop, we aimed not only to get the students interested in
Brownian motion and the frontiers of scientific research,
but also to provide an opportunity for the students to trace the fundamental research process:
setting a question, doing quantitative analysis on observed phenomena,
building a hypothesis, and verifying the hypothesis using a simple model,
even though their knowledge of mathematics and physics was limited
for understanding the mechanism of Brownian motion at the beginning of the workshop.
We think that the aims of our workshop have much in common with that of introductory physics.
Therefore, the contents presented in this article will be widely applicable to the education of
both introductory physics for life science\cite{moore_2014, mochrie_2016}
and scientific research process for the students
in the introductory-level physics course in high schools, colleges, and universities.

\section{Analysis of Brownian motion}
\begin{figure}
	\includegraphics{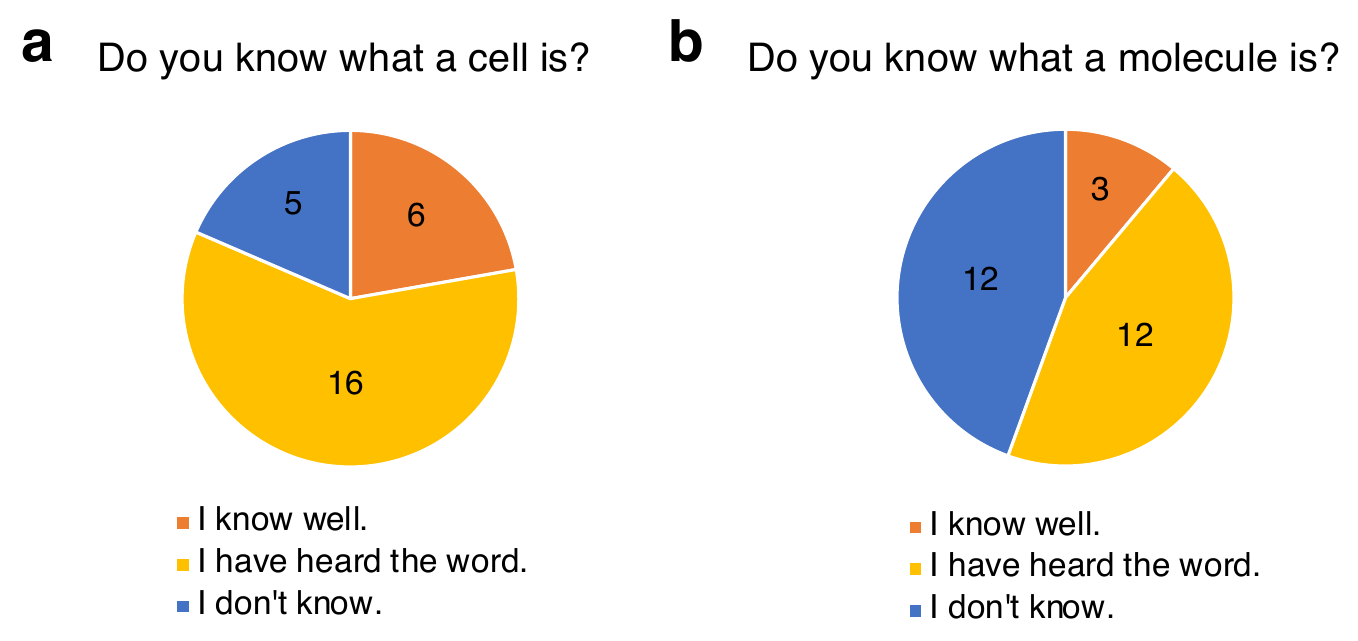}
	\caption{Results of questionnaire survey conducted before the workshop.
	Elementary school students answered two questions:
	(a) Do you know what a cell is?
	(b) Do you know what a molecule is?
	Numbers in the pie charts indicate the number of responders ($n = 27$ students).}
	\label{fig_q_bfr}
\end{figure}

Before the workshop, we administered a questionnaire to the participants.
Therein, we asked the participants whether they like science,
to which 85\% of the participants (23/27) answered in the affirmative.
Although this value was lower than the reported value in the TIMSS2019 (Japan: 92\%),\cite{timss}
participation in this workshop was voluntary, and the students appeared to be highly motivated.
We also asked them whether and how much they knew about a cell and a molecule.
The results of the two questionnaires are summarized in \figref{fig_q_bfr}.
According to the questionnaire results (\figref{fig_q_bfr}a), only 22\% of the students (6/27) knew about a cell well.
Therefore, using microscopic images, we explained at the beginning of the workshop that all living systems,
including humans, are made up of cells.
We then conducted quizzes using the voting function of the Zoom application
for the participants to recognize how small the cell is and what the cell is made of.
This is because elementary school students do not learn the units of micrometers
or the components of cells at school.
Subsequently, we introduced the unit of micrometer
and explained that typical cells are approximately about \SI{10}{\micro m} in diameter,
which is about 10 times smaller than the diameter of a single strand of hair.
We also explained that a cell is mainly composed of proteins and water,
and many kinds of tiny ``protein robots'' work ``in the pool of water'' inside the cell to maintain our life.
Note that we used the term ``protein robots'' instead of ``protein molecules'' at this stage
because elementary school students do not learn about molecules at school.
According to the questionnaire results, less than 4\% of the students (2/27) knew about molecules
(\figref{fig_q_bfr}b).

\begin{figure}
	\includegraphics[width=.5\linewidth]{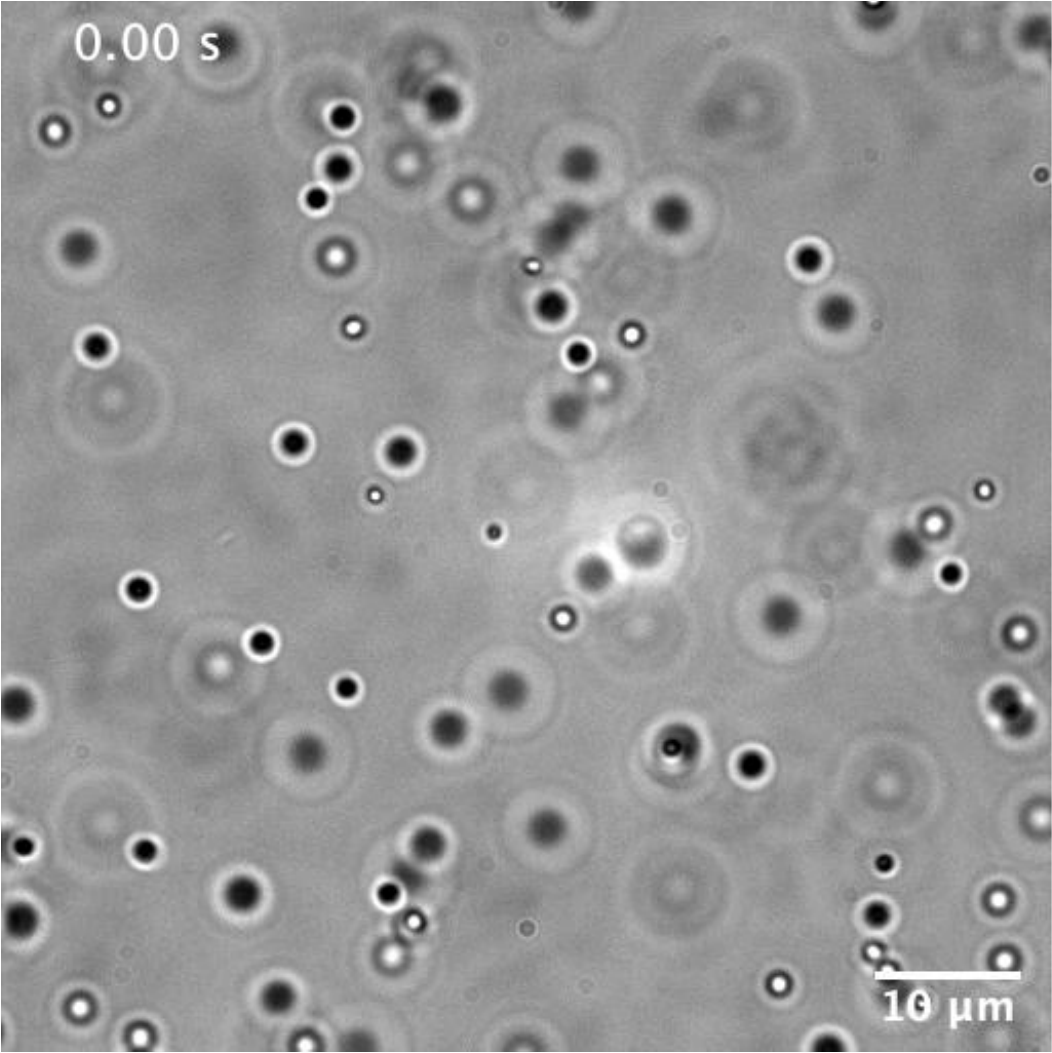}
	\caption{Snapshot from the movie of Brownian particles.
	The motion of plastic particles with a diameter of \SI{1}{\micro m} suspended in water
	was recorded using an optical microscope equipped with a $100\times$ objective lens
	and a CCD camera at a frame rate of 25 fps.
	See also \textbf{Supplementary Movie~1}.
	Additional movies for the statistical analysis are available
	from the corresponding author upon reasonable request.}
	\label{fig_brown_particles}
\end{figure}

To help the participants imagine the working environment of the protein robots in a cell,
we showed them a movie of plastic particles smaller than
the size of typical cells suspended in water (\figref{fig_brown_particles} and Supplementary Movie 1).
The particles exhibited random motion, which is generally known as Brownian motion.\cite{howard_random}
To learn the basic features of motion and infer the motile mechanism,
we asked them to analyze the motion using their personal computers at home.
We adopted the open-source software, Tracker,
as the motion analysis tool.\cite{tracker, douglas_2011}
Prior to conducting the workshop,
we compiled the original movie file in a format compatible
with the motion analysis software to simplify the analysis processes,
and we sent it to the students' parents.
Notably, we sent different movies to each family to avoid analysis of the same set of particles.
We confirmed with the parents via e-mail that the installation of the analysis software
on their personal computers was completed before the workshop.

\begin{figure}
	\includegraphics[width=\linewidth]{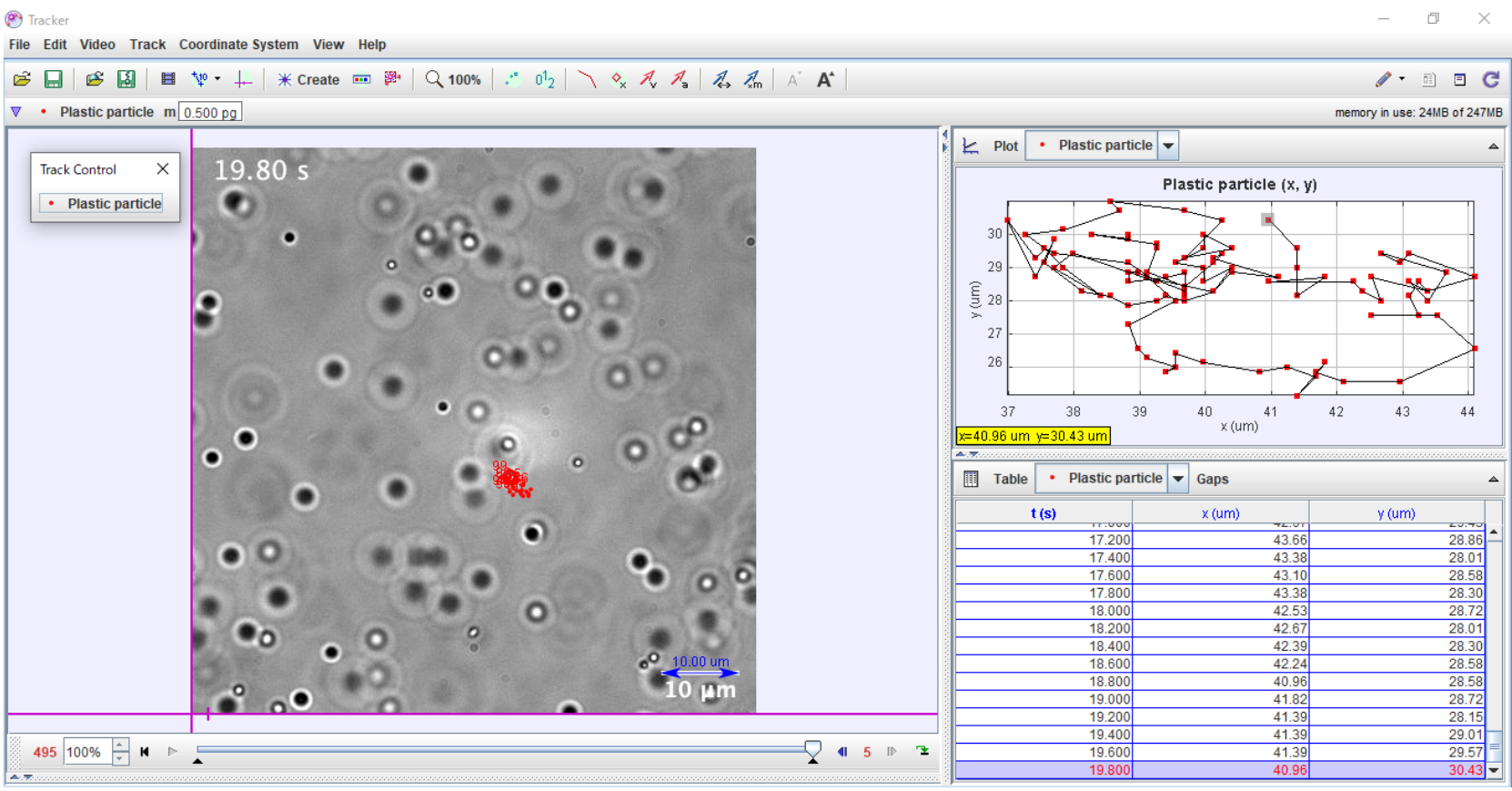}
	\caption{Screenshot of the video analysis software.
	An open-source application, Tracker, was used.
	Blurred circles are the images of defocused beads due to the motion in depth.
	We did not give the students an instruction about dealing with the Z-axis motion before the video analysis.
	When the students asked a question about defocusing, the teaching assistants taught them
	the meaning of defocusing effect and told them to keep tracking the center of the bead
	even if it was defocused.}
	\label{fig_tracker}
\end{figure}

During the practice, the students were divided into subgroups of two to six people,
and each subgroup was supported by two teaching assistants via online chat.
Using the motion analysis software,
the students recorded the positions of the particle by clicking the center every 0.20 s (every 5 frames)
for 20 s (100 time points in total) (\figref{fig_tracker}, left).
According to the Japanese math curriculum, elementary school students do not learn
about the concept of negative numbers.\cite{shido_elem}
Therefore, we set the origin at the lower left of the image
so that negative numbers did not appear in the table of X-Y coordinates (\figref{fig_tracker}, right bottom).
In addition, elementary school students had not yet learned the concept of coordinates.
We explained that the X and Y positions shown in the table indicate
how far the particle was displaced in the horizontal and vertical directions from the origin, respectively.

After the tracking stage was completed,
we asked the students to share their analysis results (a time series of X and Y positions)
using a Google spreadsheet with the help of their parents.
We carefully checked the uploaded data, and if the tracking was incorrect,
we requested the students to analyze it again, or excluded the data for further statistical analysis.

\begin{figure}
	\includegraphics[width=\textwidth]{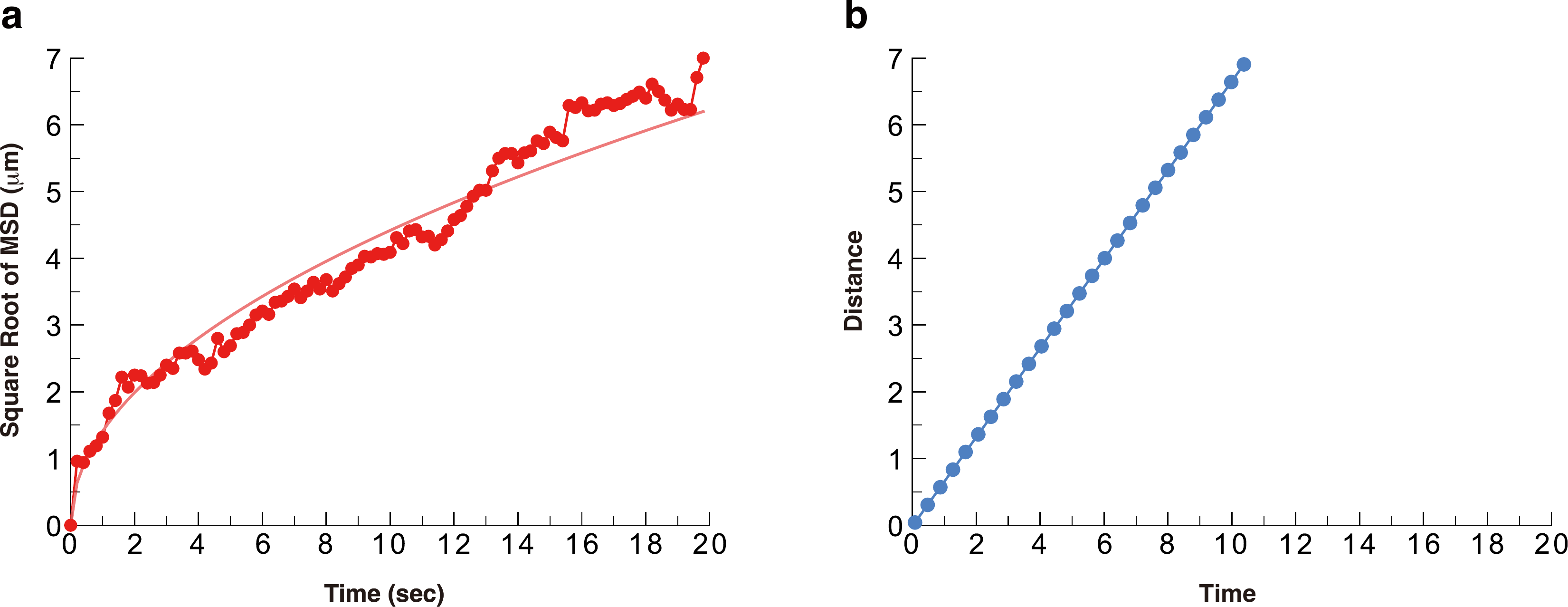}
	\caption{Results of analysis of Brownian particles from the second workshop and
	comparison with the motion of motile objects in daily life.
	(a) The relationship between time and the square root of mean-square-displacement (MSD)
	calculated from 20 tracking points is shown. Scatter plots:
	raw data, solid line: model fitting.
	(b) Conceptual graph of a car traveling at a nearly constant speed.}
	\label{fig_msd_particles}
\end{figure}

Through a comparison of the motion of motile objects in daily life, such as a car,
we discussed specific features of the motion of the analyzed particles with the students.
First, the particles moved randomly---that is, the direction of the motion changed in every video frame.
This feature is completely different from the features of regular motile objects
that such students are familiar with.
Second, the particles took four times as long to move twice as far as read from the graph.
Theoretically, it is predicted that the mean-square displacement of the analyzed particles
is proportional to the time $t$\cite{howard_random}:
\begin{equation}
	\ave{(x(t) - x(0))^2 + (y(t) - y(0))^2} = 4Dt,
	\label{eq_MSD}
\end{equation}
where $x(t)$ and $y(t)$ are the X and Y positions of a particle at time $t$ respectively,
$\ave{\,\cdot\,}$ represents the ensemble average,
and $D$ is the diffusion coefficient of the particles.
This equation gives us
\begin{equation}
	\ave{(x(t) - x(0))^2 + (y(t) - y(0))^2}^{1/2} \propto t^{1/2}.
	\label{eq_sqMSD}
\end{equation}
We confirmed that the analysis results (\figref{fig_msd_particles}a, scatter plots)
follow \eqref{eq_sqMSD} (\figref{fig_msd_particles}a, solid curve).
Here, it was expected that these equations, as well as the concept of square roots,
were difficult for elementary school students to understand.
Hence, we explained the analysis result as follows:
While motile objects in daily life such as a car can travel twice the distance in twice the time (\figref{fig_msd_particles}b),
these types of particles take four times as long to move twice as far (\figref{fig_msd_particles}a),
which is a prominent feature of the observed particles.
Based on the analysis results, it can be imagined that the tiny protein robots working
in the cell may also move at random, similar to the observed particles.

\section{2D-random walk experiment using dice}
To consider the mechanism of random motion,
we compared two movies showing the motion of plastic particles in the presence and absence of water.
This comparison clarified that water is essential for random motion.
This begs the question: Why does water induce random motion?
We proposed to the students to consider that all visible objects in nature are made up of tiny particles,
with several examples:
the human body is composed of billions of tiny cells,
a blackboard chalk can be broken down into powder when used,
and a stone is finally broken down into sand.
In other words, both chalk and stone are composed of small particles.
Using this analogy, we proposed that water would also be composed of tiny invisible particles.

\begin{figure}
	\includegraphics{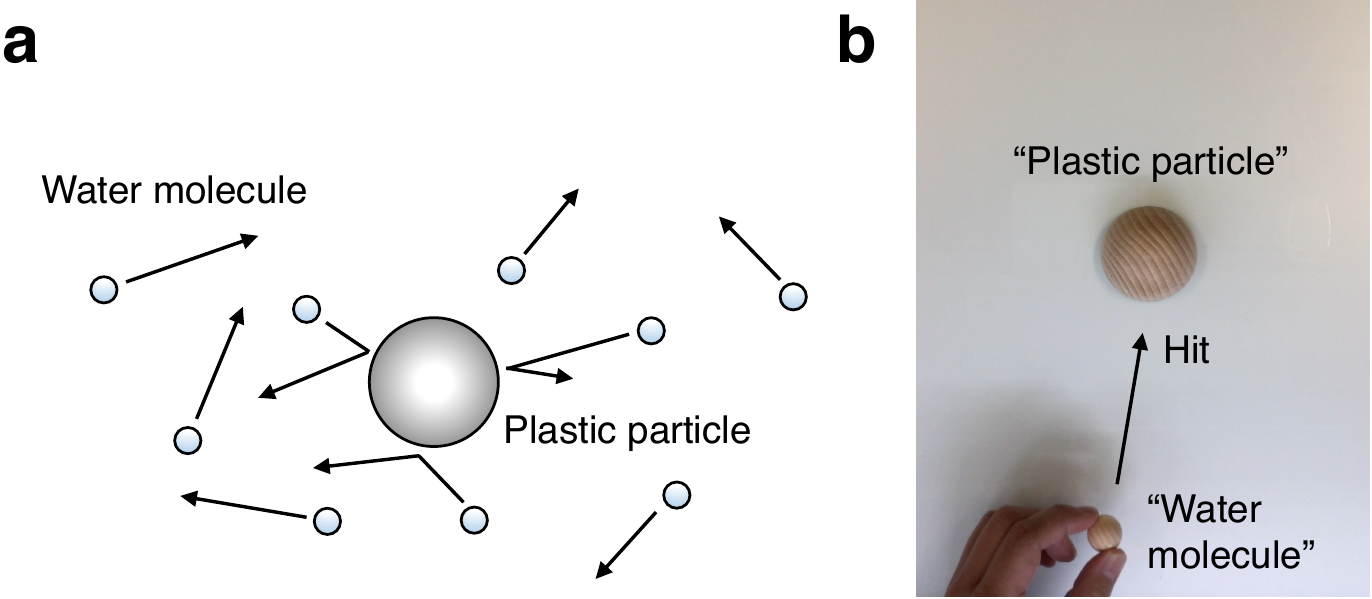}
	\caption{Mechanism of Brownian motion.
	(a) Schematic of the generally accepted mechanism. The illustration used in the lecture is shown.
	(b) Demonstration of the mechanism of Brownian motion.
	The collision effect between plastic particles and water molecules was simulated using building blocks made of wood.
	See also \textbf{Supplementary Movie 2}.}
	\label{fig_mechanism}
\end{figure}

\begin{figure}
	\includegraphics[width=\textwidth]{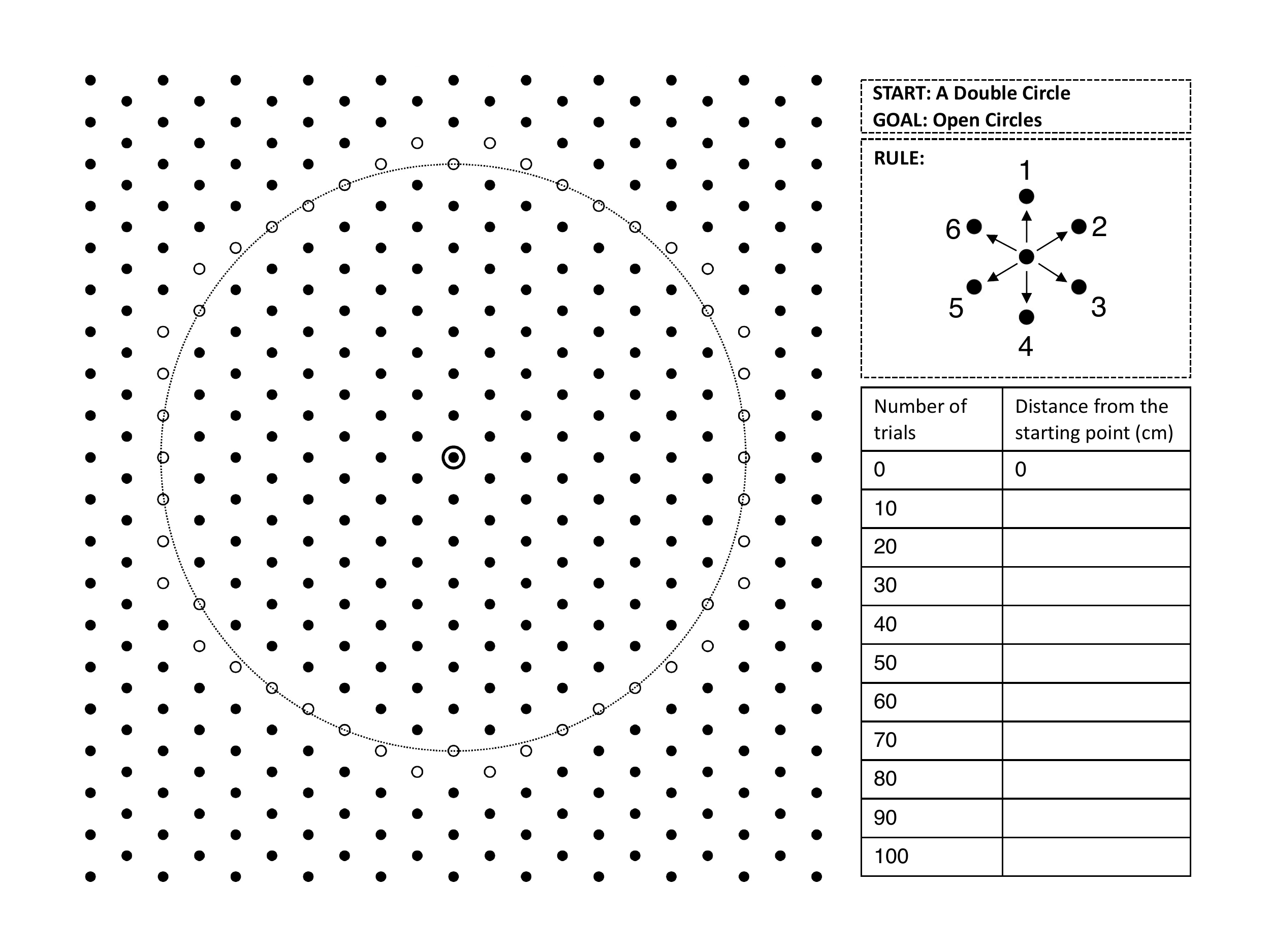}
	\caption{Dice game sheet for simulating 2D-random walk,
	printed on A4-sized papers and sent to the participants prior to the workshop.
	The original game sheet is available in \textbf{Supplementary Sheet}.}
	\label{fig_dice_sheet}
\end{figure}

\begin{figure}
	\includegraphics{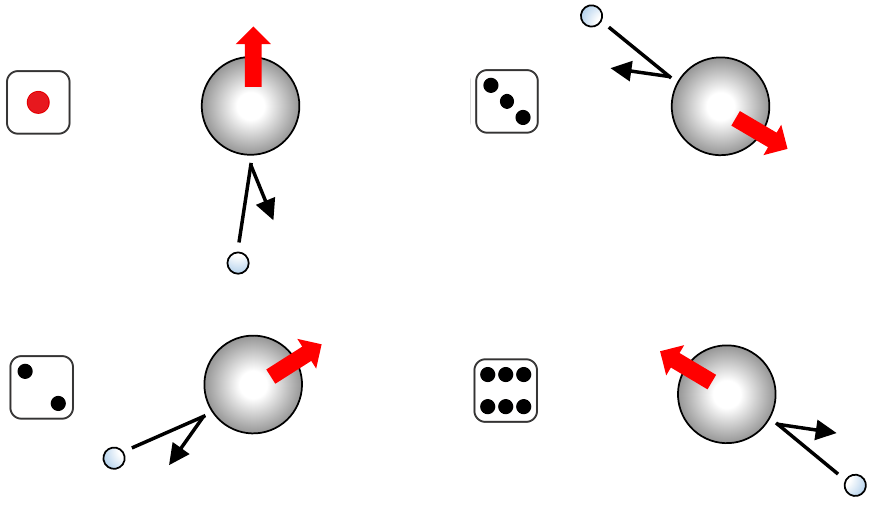}
	\caption{Rolling dice can emulate the random collision events of water molecules with a plastic particle.
	An illustration used in the lecture is shown.}
	\label{fig_dice_collision}
\end{figure}

Thereafter, we introduced the generally accepted mechanism of Brownian motion:
Collisions of water molecules push the plastic particles at random (\figref{fig_mechanism}a),
as a hypothesis, not as knowledge.
To help clarify the collision effect, we used building blocks composed of wood to demonstrate that
the hypothesis could possibly be correct (\figref{fig_mechanism}b and Supplementary Movie 2).
To verify this hypothesis, we performed a hands-on experiment
using a dice game sheet\cite{oosawa_hand} (\figref{fig_dice_sheet} and Supplementary Sheet).
The sheet was printed on A4-sized papers and then sent to the participants prior to the workshop.
The starting point was set at the center, indicated by a double circle (\figref{fig_dice_sheet}, left).
The direction of movement was determined by the roll of the dice (\figref{fig_dice_sheet}, top right).
This rule emulated random collision events of water molecules to the particles (\figref{fig_dice_collision}).

The students rolled the dice and moved a token around the game sheet
according to the rules shown in \figref{fig_dice_sheet}.
Every 10 trials, they measured the distance between the current position of the token
and the starting point with a ruler and entered the value into a spreadsheet (\figref{fig_dice_sheet}, bottom right).
We set the goal points at positions seven steps apart from the starting point,
as indicated by the open circles, to include a gamified environment.
During the experiments, the students were divided into subgroups and supported by the teaching assistants.

\begin{figure}
	\includegraphics[width=.6\textwidth]{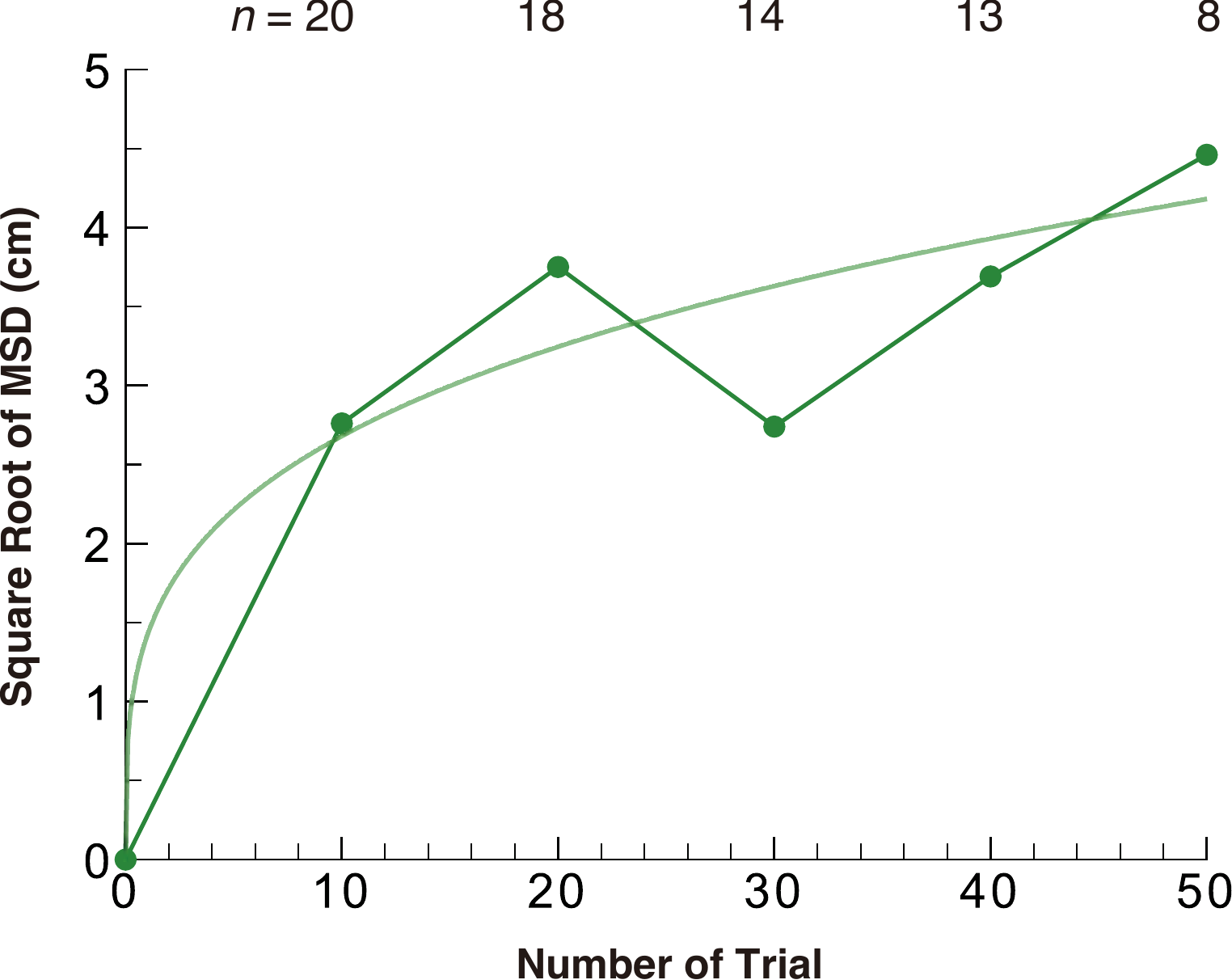}
	\caption{Results of analysis of 2D-random walk experiment using dice, from the second workshop.
	The relationship between the number of trials and the square root of mean-square-displacement (MSD)
	calculated from 20 independent experiments is shown.
	Scatter plots: raw data, solid line: model fitting, n: sample numbers used for the calculation of MSD in each plot.}
	\label{fig_msd_dice}
\end{figure}

The results shown in \figref{fig_msd_dice} show a similar trend to
that of the plastic particles (\figref{fig_msd_particles}).
In particular, the displacement does not seem proportional to the number of rolling dice $N$,
but is rather proportional to $N^{1/2}$ (equivalent to $t^{1/2}$ in the case of plastic particles).
We explained to the students that our hypothesis was correct,
and thus, the mechanism of Brownian motion was understood.

\section{Bridging the motion of plastic particles with protein dynamics in a cell}
\label{sec:bridging}

In the last section, we introduced the unit of nanometer and explained that
any objects in nature are composed of nanoscale tiny particles, namely molecules.
In a cell, protein molecules are exposed to the random collisions of water molecules
and thus they exhibit Brownian motion.
Based on the results of the video analysis and dice experiment,
we discussed that protein molecules will move a short distance quickly,
but it will take a long time to travel a long distance solely by the Brownian motion.
We then introduced the ``special protin robot'' called molecular motor exists in the cell.
The molecule motor transports the other protins rapidly and unidirectionally along the tracks
made of the other proteins, just like a car or train in the world of humans.
We then showed some animations\cite{kinesin_mov}
and microscopic images of molecular motors such as kinesin and myosin\cite{kodera_2010}
walking along the cytoskeletal networks.
Growing evidence obtained from recent single-molecule experiments
and advances in the theory of information statistical mechanics suggest that
molecular motors utilize collision of water molecules as the major driving force
for the unidirectional motion,
and the energy supplied by ATP hydrolysis is used to bias the direction of the stepping motion.
\cite{busta_2001, shiroguchi_2007, sagawa_2010, toyabe_2010}
Supported by these findings, we demonstrated using animations
that molecular motors utilize collisions of water molecules as the driving force
by converting the random force into unidirectional motion efficiently.
State-of-the-art technologies, including the development of synthetic ``molecular engines,''
as well as future visions of molecular sciences, were also discussed.\cite{iino_2020, toyabe_2020, kinbara_2021}

\section{Conclusions and Discussion}
\label{sec:conclusions}

\begin{figure}
	\includegraphics{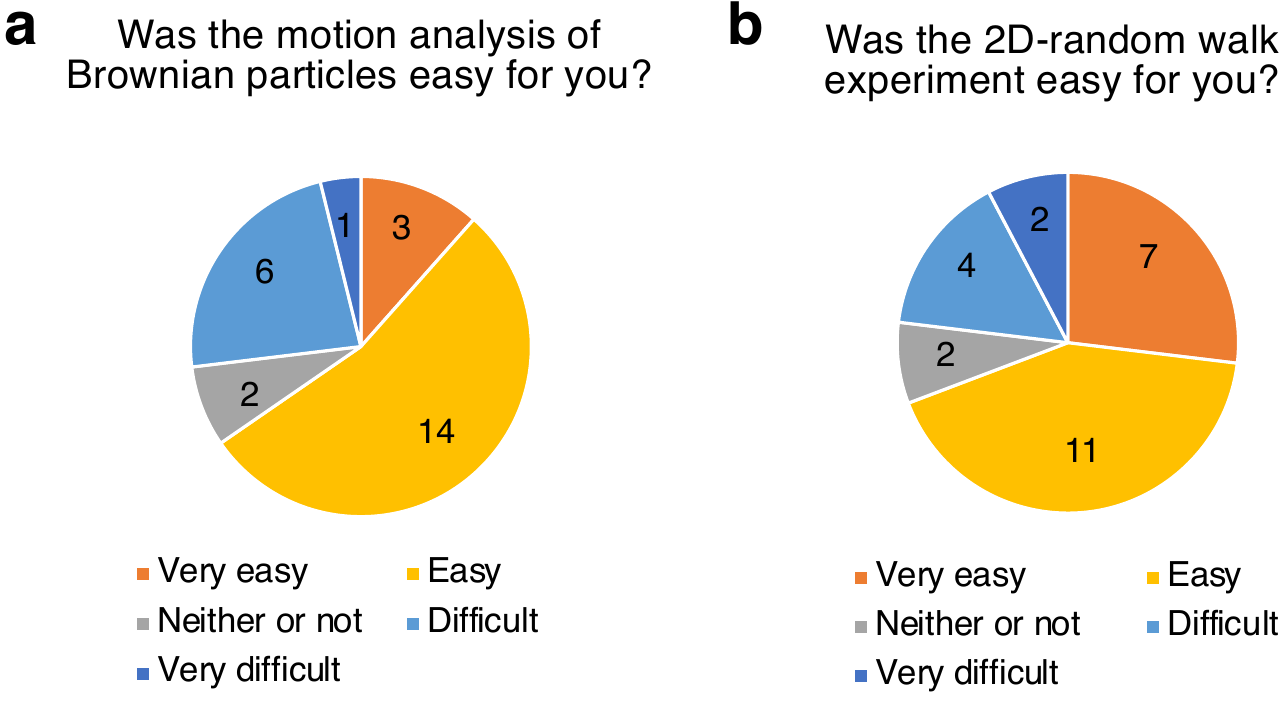}
	\caption{Results of questionnaire survey conducted after the workshop.
	Elementary school students answered two questions:
	(a) Was the motion analysis of Brownian particles easy for you?
	(b) Was the 2D-random walk experiment easy for you?
	Numbers in the pie charts indicate the number of responders ($n' = 26$ students).}
	\label{fig_q_aft}
\end{figure}

After the workshop, we conducted a questionnaire-based survey.
\figref{fig_q_aft}a shows whether the video analysis was easy for the students to perform.
More than half of the respondents replied that the video analysis was very easy or easy (17/26).
We believe that difficulties in performing such video analyses are
largely dependent on the complexity of computer operations,
such as importing raw data files and setting the length scale and time intervals.
In this workshop, we prepared a compiled analysis file ready for particle tracking just by mouse clicking.
We successfully demonstrated that even elementary school students can work on video analysis
using personal computers without much stress, if the initial analysis file is prepared properly.

\figref{fig_q_aft}b shows whether the 2D-random walk experiment was easy for the students to perform.
More than half of the respondents replied that the experiment was very easy or easy (18/26).
As we set a goal to include a gamified environment,
many students were excited for and enjoyed this experiment.
However, we realized that the sample number $n$ decreased
as the number of trials $N$ increased (see \figref{fig_msd_dice}).
This was because once a token reached the goal, the data for larger $N$ values was lost.
To improve the data accuracy, it would be better not to set a goal in this experiment.

We found that our workshop has a limit on its applications.
Weak internet connection sometimes causes a delay on the time schedule both in the lecture and practice parts.
In addition, video analysis requires that all the participants have a personal computer.
It is important to study how we can arrange the workshop program for widely applicable package
in various situations.

\begin{figure}
	\includegraphics{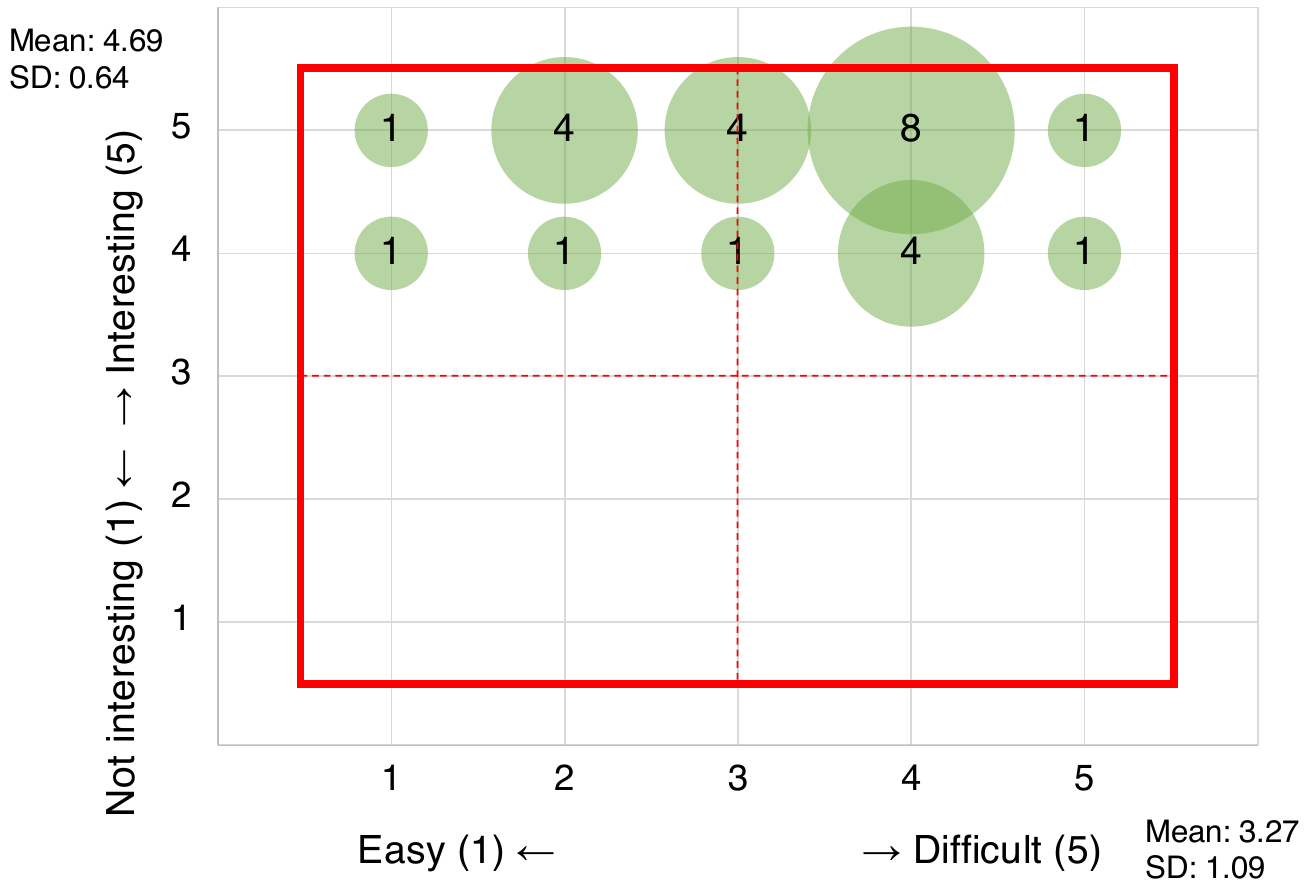}
	\caption{Bubble chart showing questionnaire results obtained after the workshop.
	We asked the participants how interesting and difficult the workshop was, on a scale of 1 to 5 each.
	Numbers in each bubble indicate the number of responders.
	Mean values and the standard deviations (SDs) of five-grade evaluations
	for each questionnaire are also displayed in the figure ($n' = 26$ students).}
	\label{fig_bubble_chart}
\end{figure}

Although we were forced to conduct the workshop in the form of distance education
because of the COVID-19 pandemic,
we received positive responses from all the respondents.
Interestingly, more than 50\% of the respondents (14/26) replied
that the workshop was very difficult (score 5; 2/26) or difficult (score 4; 12/26) to understand,
but all the respondents replied that the workshop was very interesting (score 5; 18/26)
or interesting (score 4; 8/26) (\figref{fig_bubble_chart}).
This result indicates that motivated students were fascinated with the experimental learning and the lectures,
even though the contents may have been difficult to grasp.
Indeed, as per the Japanese curriculum, Brownian motion is taken up in a topic
of the thermal fluctuation of atoms and molecules only in high school
or at the university education level.
The typical situation is that students simply observe the random motion of small particles
under an optical microscope,
then they are taught by teacher that the motion reflects
the thermal fluctuation of atoms or molecules around the particle.
In contrast, in our workshop, the students learned the mechanism of Brownian motion
by tracing the fundamental research process:
setting a problem, doing quantitative analysis, building a hypothesis, and verifying the hypothesis
using a simple model.
By devising the contents of experiments and lectures, even elementary school students
who did not have advanced knowledge of physics and mathematics began to take an interest in Brownian motion.
In summary, we found that ``difficult but interesting'' workshops,
in which the contents are systematically organized for beginners,
are preferable to stimulate the intellectual curiosity of motivated students
in the early stage of scientific education.

\section{Acknowledgements}
We thank Ken'ya Furuta for providing movies of kinesin molecules walking along microtubules
captured by TIRF microscopy,
Haruka Iwai, Natsumi Ito, Rito Ikeda, Tatsuya Kagemoto, Kosuke Nakamura, Ami Shiraishi,
Kaoru Shinoda, and Yukie Shirota for their assistance in the image analysis and dice experiments.
This work was partly supported by the Tokyo Institute of Technology Fund (grant no.~T130SH3152 to Y.H.),
Scientific Research on Innovative Areas ``Molecular Engines'' (grant nos.~19H05393 and 21H00399 to M.M.),
JST PRESTO (grant no.~JPMJPR20ED to M.M.), and the Hakubi Project of Kyoto University (to M.M.).

\clearpage

\clearpage

\clearpage

\clearpage

\clearpage

\clearpage

\clearpage

\clearpage

\clearpage

\clearpage

\end{document}